# Fibonacci terahertz imaging by silicon diffractive optics

D. Jokubauskis*, L. Minkevičius, M. Karaliūnas, S. Indrišiūnas,
I. Kašalynas, G. Račiukaitis, G. Valušis

*Center for Physical Sciences and Technology, Saulėtekio Ave. 3, LT-10257 Vilnius, Lithuania*
*Corresponding author: domas.jokubauskis@ftmc.lt,*

**Fibonacci or bifocal terahertz (THz) imaging is demonstrated experimentally employing silicon diffractive zone plate (SDZP) in a continuous wave mode. Images simultaneously recorded in two different planes are exhibited at 0.6 THz frequency with the spatial resolution of wavelength. Multi-focus imaging operation of the Fibonacci lens is compared with a performance of the conventional silicon phase zone plate. Spatial profiles and focal depth features are discussed varying the frequency from 0.3 THz to 0.6 THz. Good agreement between experimental results and simulation data is revealed.**

https://doi.org/10.1364/OL.43.002795

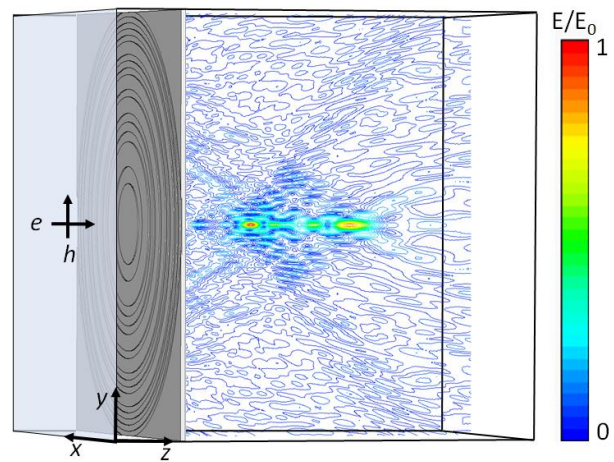

**Fig. 1.** The normalized electric field ratio $E/E_0$ distribution in x and z directions before and after the Fibonacci diffractive zone plate at 0.6 THz frequency. Data are obtained using 3D Finite-difference time-domain (FDTD) simulations at the depicted coordinate axes, electric and magnetic field orientations. Z axis shows the THz beam propagation direction. Note that the distribution of the electric field exhibits extended focusing performance at a several points along the beam propagation path.

Terahertz (THz) imaging and spectroscopy displays an attractive feature to expose new possibilities and wide-ranging potential in many versatile applications. Already known implementations covering security [1,2], package inspection [3], biomedicine diagnostics [4,5] and food control [6] recently have been extended by promising studies in investigating 2D materials [7,8], surface control [9] and paintings examination [10,11]. From practical point of view, design of a THz imaging system should be compact, free of optical alignment, reliable and providing an ability of relatively rapid scans. Moreover, significant preference could be not only to record a structure of the sample, but, via determination of spectroscopic features, identify the object and, if possible, to reconstruct its three-dimensional image. Interesting and elegant approach to combine advantages of compact diffractive optics with benefits of tomographic imaging was given in [12,13]. Employing terahertz time-domain spectroscopy (THz-TDS) and unique feature of Fresnel lens which focal length is linearly proportional to the frequency, tomographic images of a target using multiple frequencies were recorded [14]. In other words, operating in a pulsed mode and utilising a Fresnel lens at different frequencies of the imaging beam allows objects reconstruction at various positions along the beam propagation path onto the same imaging plane.

In this paper, we convert the aforementioned concept into the continuous wave and discrete frequency operation mode via engagement of Fibonacci diffractive lenses [15]. In contrast to recently presented 3D printed diffractive terahertz lenses [16], we demonstrate design, operation and high-resolution imaging of silicon-based Fibonacci lens for 0.6 THz frequency. It was fabricated on a monocrystalline silicon wafer using laser patterning earlier employed to produce multilevel phase Fresnel lenses of high efficiency [17]. The focusing performance was investigated theoretically and experimentally by measuring spatial profiles, the distance between the foci and focal depth at 0.3 THz and 0.6 THz. The ability to perform simultaneous imaging with the wavelength resolution of two planes separated by 7 mm distance was experimentally revealed. The multifocal imaging results were compared with the performance of the phase zone plate designed of the same diameter and material.

The Fibonacci lens design relies on Fibonacci sequence principle when the next number of the sequence is obtained adding up the two preceding ones. Based on the Fibonacci numbers, a binary aperiodic Fibonacci structure can be composed indicating that each element of the sequence is found as the concatenation of the two previous elements. As a result, non-uniform mapping of this structure provides the radial profile of the Fibonacci lens [15].

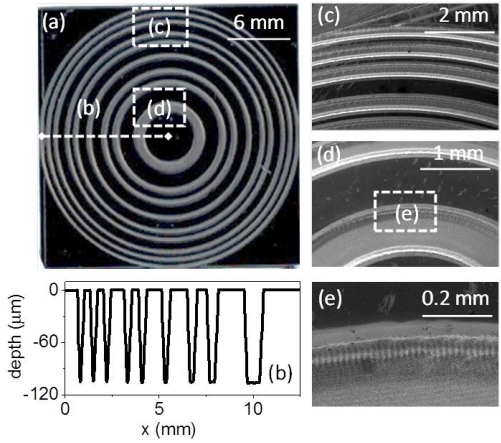

**Fig. 2.** (a) SEM images of the 0.6 THz Fibonacci (bifocal) silicon diffractive zone plate with the focal distances of f = 7 mm and 13 mm; (b) the cross-section profile of ablated grooves in silicon surface; (c) and (d) zoomed area of outer and center rings, respectively; (e) displays the junction between the ablated and polished silicon surface.

In our approach, silicon, which served as a highly efficient material for multilevel phase Fresnel lenses [17], was chosen for the Fibonacci (bifocal) lens production. Initially, theoretical modelling of the electric field distribution in a space after the bifocal diffractive silicon zone plate was performed by using three-dimensional finite-difference time-domain (3D FDTD) method. The spatial resolution was set-up from 0.01 mm to 0.115 mm. To simplify the simulation, symmetry conditions of the structure were applied, and the absorbing boundary conditions were set in all the directions. The normalised electric field was recorded in the whole simulation volume. Illustrative simulation results for the volume of 22×22×15 $mm^3$ in $x, y, z$ directions are presented in Fig. 1. A source of a multi-frequency plane wave, transparent to the reflected wave, was specified in the front of the zone plate plane (grey semi-transparent rectangle in Fig. 1). The predicted focusing performance of the Fibonacci lens is demonstrated in Fig. 1. As one can see the distribution of the electric field exhibits extended focusing performance at several points along the beam propagation with maximal intensity in the two separate focal spots of the lens. Note that the peak of the first amplitude in the z-direction is higher than the second one. Hence, such a lens can be a reasonable instrument to enable a bifocal regime of operation and provide, therefore, an option to record simultaneously images of objects placed at the different foci.

One can underline that the focusing picture is colored by a pattern of standing waves observed in reflection due to the incident and reflected plane wave interference.

According to the simulation results, the design of the silicon Fibonacci lens was developed. The focusing element was fabricated using laser ablation technology based on industrial-scale laser-direct-write (LDW) system based on a 1064 nm wavelength, a 13 ps pulse, 1 MHz repetition rate, and a 60 µJ peak energy laser (Atlantic 60 from Ekspla Ltd.) in conditions described precisely in [17]. Sample with outside diameter of 24 mm and two foci separated by a defined distance of 7 mm was fabricated on monocrystalline silicon (orientation (110), with resistance 0.01-1 MΩcm and refractive index 3.46. The SEM images of the lens are shown in Fig. 2. The step-profile scanned across the center of the sample and enlarged zones of the ablated rings are displayed in Fig. 2(b-e). It is seen that the ablated grooves near the center and near the edge of the diffractive element are ablated to the depth of 100 µm.

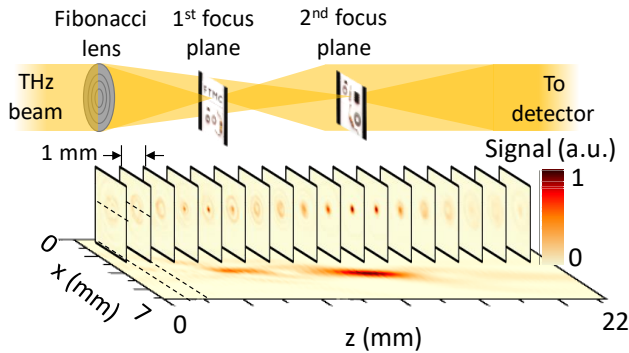

**Fig. 3.** Experimental demonstration of 0.6 THz radiation beam profiles focusing evolution along the beam propagation path. The cross-sections represent the beam alternation in focal plane direction. The distance between cross-sections - 1 mm. 2D images consist of 66 x 67 pixels. Pixel size 0.1 x 0.1 $mm^2$. The dashed line represents the cross-section position in xz plane. Upper panel depicts part of the experimental set-up to record Fibonacci imaging.

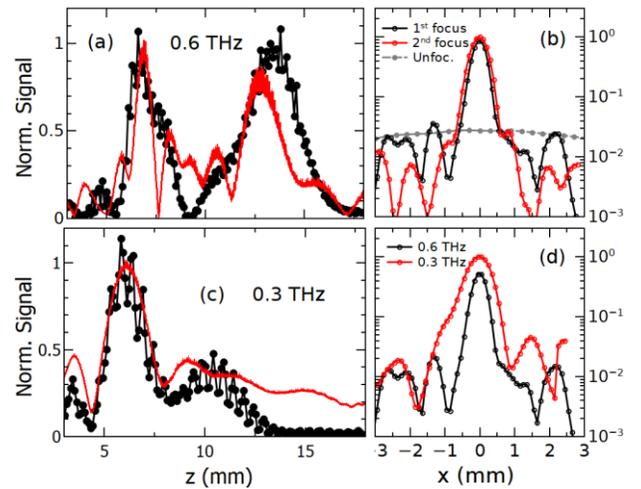

**Fig. 4.** Normalized THz radiation power distribution along z-axis (a) and along x-axis at the 1st and 2nd focus (b) measured for 0.6 THz frequency radiation. The unfocused relatively scaled 0.6 THz radiation profile is depicted for comparison. The THz power distribution for 0.3 THz is depicted along z-axis (c) and x-axis (d) where it is compared with the distribution along x-axis for 0.6 THz measured at the 1st focus. Red solid lines in panels (a) and (c) are the results of simulation using 3D FDTD method.

The focusing performance of the bifocal silicon diffractive zone plate was investigated by measuring Gaussian beam intensity distribution in the focal plane and along the optical axis at the 0.3 THz and 0.6 THz frequencies. Experimental demonstration of the focusing was evidenced using the experimental setup described in [17]. The radiation was collimated by 12 cm focal length high-density polyethylene (HDPE) lens and registered with resonant THz antenna-coupled micro-bolometer detector [4]. The intensity of THz radiation distribution along the beam propagation direction and cross-sections at $z$-direction with a step of 1 mm was measured. The results of the focusing performance are presented in Fig. 3. It is seen that two clearly distinguished focal spots are observed at different distances, 6 mm and 13 mm, from the Fibonacci focusing element. These foci are located at certain axial positions determined by the Fibonacci numbers [15]. It is worth noting that noticeable Airy disks are pronounced between the first and second focal spots.

Aiming to prove that indeed the observed foci are related to the Fibonacci lens performance, the cross-section of the measured profiles and simulated focusing operation was compared (Fig. 4.) The distribution of THz radiation power for 0.6 THz along the $z$-axis is depicted in panels (a), (b) and (d). For more illustrative comparison, additionally, the data obtained at 0.3 THz (panel (c)) are presented as well. It is evident that two well-resolved foci can be identified at 6.7 and 13.4 mm away from the Fibonacci lens surface when illuminating the lens with the 0.6 THz frequency radiation. One can note that experimental data fit well with the simulated results. Moreover, the first maximum along the $z$-axis is narrower than the second one which is also in accordance with the simulated data obtained by FDTD. The beam profile in the $x$-direction exhibits similar features: the first focus is also narrower than the second one with the corresponding full-width half-maximum (FWHM) values of 0.46 and 0.55 mm, respectively. Note that the FWHM of the unfocused Gaussian beam before the Fibonacci lens reaches 8.55 mm, and the signal amplitude is nearly two orders of magnitude lower as it can be nicely seen in Fig 4 (b).

To get a more expressive picture via comparison, the radiation frequency was detuned to 0.3 THz, then both focusing operation and profile features were evaluated and compared with that at 0.6 THz. It is seen from Fig. 4(c) that experimental results precisely follows the simulation data: the second focus becomes smooth and demonstrates a trend to disappear, while the 1st one experiences no significant change. If one looks from the profile point of view, in the $x$-axis direction, the signal amplitude at 0.3 THz is in a factor of 2 higher than that at 0.6 THz at the first focus as it is compared in Fig. 4 (d). The FWHM along the $x$-axis for 0.3 THz radiation is 0.76 mm, while for 0.6 THz it amounts to 0.46 mm.

As the bifocal operation is clearly resolved, one needs to evaluate its spatial resolution and demonstrate the THz Fibonacci imaging. Spatial resolution was investigated using specially constructed resolution target consisting of periodic metallic stripes (photo depicted in Fig. 5 (a)) with different distances in between. As one can see, the smallest period of stripes which was clearly resolved in both foci is 0.8 mm, i. e. it amounts to 1.6 factor of wavelength, 1.6×λ. Therefore, reasonable spatial resolution can be expected and excellent quality in recording THz images in both focal planes simultaneously.

For the Fibonacci imaging aims a special sample was constructed. It is based on a plastic frame of 7 mm thickness to fit the distance between both foci as is illustrated in (Fig. 4 (a)). The frame was then covered with a sticky tape from both sides enabling thus, on the one hand, transparency for THz light and ability to record images in transmission geometry, on the other hand, to realise two planes for attachment of the objects under test. To get broader illustration different objects were chosen (Fig. 6 (a)). The sample's 1st plane was composed of "FTMC" – Lithuanian abbreviation of the Center for Physical Sciences and Technology – written in pencil on the office paper, resistor with 0.5 mm width leads, SMD 1206 resistor, 4.6 mm width washer, 4 mm width nut, 6.7 mm width washer. The sample's 2nd plane consists of SMD 1206 resistor, 4.6 mm width washer, 0.5 mm width pin, 1.27 mm pin pitch SOIC package, resistor with 0.5 mm width leads, 6.7 mm width washer. Terahertz imaging results obtained using various diffractive optics elements are shown in Fig. 6, panels (b-e). Each image was acquired in 500 s time and consists of 400 x 250 pixels with 0.3 mm pitch.

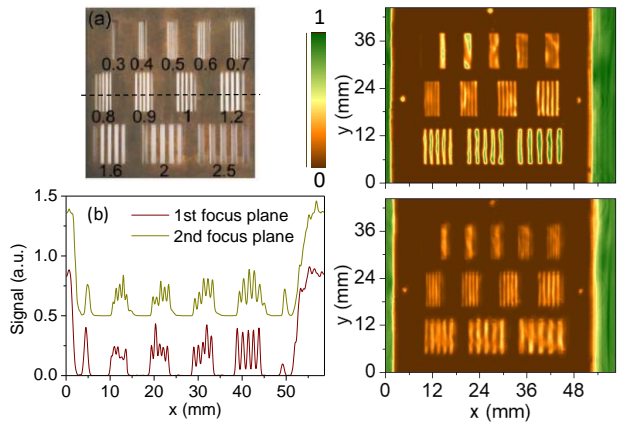

**Fig. 5.** Photo (a) of the resolution target with indications of the period in mm scale. Dashed line indicates measuring place to determine cross section; (b) – cross section of the THz beam profile along the x-axis at y=22 mm, data lines shifted by a factor of 0.5 for convenience of illustration. THz images of resolution target at 0.6 THz: (c) panel presents image at 6.6 mm away from the Fibonacci lens (the first focus plane) while (d) shows – at 13.3 mm (the second focus plane).

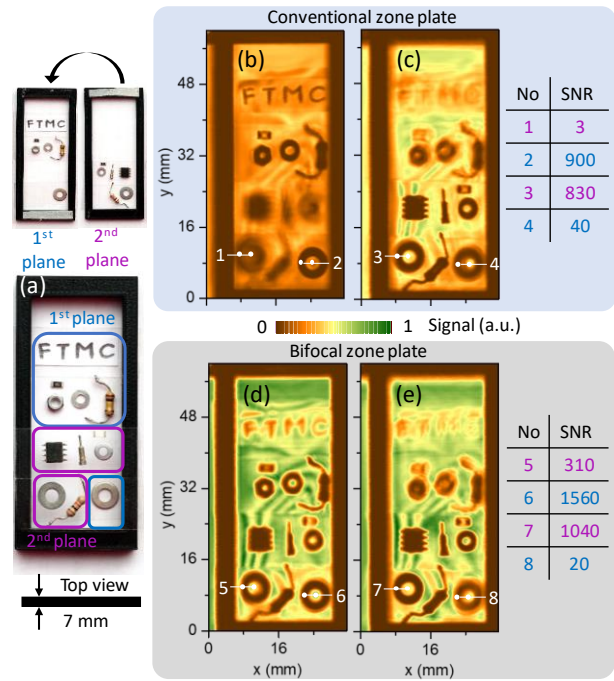

**Fig. 6.** THz images at two planes, separated by 7 mm spacer obtained simultaneous recording at 0.6 THz. Photos of the sample (a). THz images at 0.6 THz obtained using conventional zone plate: image focused on the 1st plane (b) and focused on the 2nd plane (c). Images obtained using bifocal zone plate: optimal image taken by moving the sample along $z$-axis to achieve the best contrast in both sample planes (d) and image taken at the highest contrast in the 2nd plane contrast (e). Positions of signal amplitudes used for SNR calculations are represented as white lines with dots.

Initially, the conventional diffractive silicon phase zone plate, SDZP, [17] with the focal distance of 10 mm was investigated, and relevant images were recorded. THz images at 0.6 THz are displayed in panels (b) and (c) of Fig. 6. Results show that using SDZP the THz images of good quality (signal-to-noise ratio, S/N, is more than 800) can be recorded aligning imaging to relevant foci. It is clearly seen from Fig. 6(b), when the focus is tuned to the first plane, it enables to image objects in the first plane, even pencil written letters "FTMC" – are evidently resolved. However, the image quality of the objects placed in the second plane is not tolerable – images are blurred, S/N ratio is in the range of 40. Similar picture was obtained with the tuning to the second plane – in that case objects placed in the second plane can be plainly distinguished (S/N ratio is in the order of 900), but the THz image quality in the first plane is strongly decreased, especially it can be seen that the letters "FTMC" are hardly to be resolved. (Fig. 6(c)). Therefore, the conventional phase zone plate is not suitable for simultaneous THz imaging in both planes.

In contrast, in bifocal zone plate case, the image was recorded collecting the THz light from both foci in a single scan. Results are depicted in Fig. 6 (d). Since both planes are in focus now, all the objects placed in both planes as well as pencil written letters "FTMC" are clearly resolved. S/N ratio in the first and the second foci is respectively in the order of 1500 and 300.

It is interesting to compare operation of both diffractive components – conventional SDZP and the Fibonacci lens. To set the same recording conditions, the imaging system was tuned for THz radiation collection just from the 2nd focus of bifocal zone plate in order to compare the resolution with the conventional SDZP. The obtained image is shown in Fig. 6 (e). If compare it with the image given in Fig. 6 (c), one can see that second sample plane spatial resolution is comparable to conventional zone plate image, while the image resolution in the first plane is strongly decreased. It let to conclude that two plane THz imaging can be performed using Fibonacci (bifocal zone) plate reaching simultaneously wavelength resolution in both foci.

To conclude, the silicon-based Fibonacci (bifocal) diffractive zone plate for 0.6 THz frequency has been developed, and its employment for bifocal THz imaging with the wavelength resolution was demonstrated. It not only enriched a family bifocal focusing elements like bifocal zoom lenses using plasmonic metasurfaces [18], fractal zone planes with reduced aberration in visible range [19], plates based on generalized $m$-bonacci sequence [20], but also extended routes to manipulate with THz beam [21] in continuous wave mode using diffractive optics components providing thus an additional tool for further development of compact and alignment-free THz imaging systems.

**Funding.** This work was supported by the Research Council of Lithuania (LAT 04/2016).

**References**

1. M. Kato, S. R. Tripathi, K. Murate, K. Imayama, and K. Kawase, Opt. Express **24**, 6425 (2016).
2. U. Puc, A. Abina, M. Rutar, A. Zidanšek, A. Jeglič, and G. Valušis, Appl. Opt. **54**, 4495 (2015).
3. L. Minkevičius, V. Tamošiunas, I. Kašalynas, D. Seliuta, G. Valušis, A. Lisauskas, S. Boppel, H. G. Roskos, and K. Köhler, Appl. Phys. Lett. **99**, 97–100 (2011).
4. I. Kašalynas, R. Venckevičius, L. Minkevičius, A. Sešek, F. Wahaia, V. Tamošiūnas, B. Voisiat, D. Seliuta, G. Valušis, A. Švigelj, and J. Trontelj, Sensors **16**, 432 (2016).
5. B. C. Q. Truong, A. J. Fitzgerald, S. Fan, and V. P. Wallace, Biomed. Opt. Express **9**, 1334 (2018).
6. A. A. Gowen, C. O'Sullivan, and C. P. O'Donnell, Trends Food Sci. Technol. **25**, 40–46 (2012).
7. J. D. Buron, F. Pizzocchero, P. U. Jepsen, D. H. Petersen, J. M. Caridad, B. S. Jessen, T. J. Booth, and P. Bøggild, Sci. Rep. **5**, 1–7 (2015).
8. P. R. Whelan, K. Iwaszczuk, R. Wang, S. Hofmann, P. Bøggild, and P. U. Jepsen, Opt. Express **25**, 2725 (2017).
9. K. Ueno, S. Nozawa, and H. Misawa, Opt. Express **23**, 28584 (2015).
10. C. L. K. Dandolo, J.-P. Guillet, X. Ma, F. Fauquet, M. Roux, and P. Mounaix, Opt. Express **26**, 5358 (2018).
11. C. L. Koch Dandolo, M. Picollo, C. Cucci, M. Ginanni, E. Prandi, M. Scudieri, and P. U. Jepsen, J. Infrared, Millimeter, Terahertz Waves **38**, 413–424 (2017).
12. X. C. Zhang, Philos Trans. Ser A Math Phys Eng Sci **362**, 283–289 (2004).
13. S. Wang, B. Ferguson, D. Abbott, and X. C. Zhang, J. Biol. Phys. **29**, 247–256 (2003).
14. S. Wang and X.-C. Zhang, J. Phys. D. Appl. Phys. **37**, R1–R36 (2004).
15. J. A. Monsoriu, A. Calatayud, L. Remon, W. D. Furlan, G. Saavedra, and P. Andres, IEEE Photonics J. **5**, 3400106–3400106 (2013).
16. W. D. Furlan, V. Ferrando, J. A. Monsoriu, P. Zagrajek, E. Czerwińska, and M. Szustakowski, Opt. Lett. **41**, 1748 (2016).
17. L. Minkevičius, S. Indrišiūnas, R. Šniaukas, B. Voisiat, V. Janonis, V. Tamošiūnas, I. Kašalynas, G. Račiukaitis, and G. Valušis, Opt. Lett. **42**, 1875 (2017).
18. X. Yin, T. Steinle, L. Huang, T. Taubner, M. Wuttig, T. Zentgraf, and H. Giessen, Light Sci. Appl. **6**, e17016 (2017).
19. W. D. Furlan, G. Saavedra, and J. A. Monsoriu, Opt. Lett. **32**, 2109–2111 (2007).
20. F. Machado, V. Ferrando, W. D. Furlan, and J. A. Monsoriu, Opt. Express **25**, 8267 (2017).
21. L. Minkevičius, K. Madeikis, B. Voisiat, I. Kašalynas, R. Venckevičius, G. Račiukaitis, V. Tamošiunas, and G. Valušis, J. Infrared, Millimeter, Terahertz Waves **35**, 699–702 (2014).